          [2001/04/25 1.1 (PWD)]
\documentclass[a4paper,twocolumn]{esapub2005} 
\usepackage{bm}
\usepackage{url}
\usepackage{times}
\usepackage{natbib}
\usepackage{graphicx}

\title{Connection between active longitudes and magnetic helicity}
\author{Axel Brandenburg}
\affil{Nordita, Blegdamsvej 17, DK-2100 Copenhagen \O, Denmark}
\author{Petri J.\ K\"apyl\"a}
\affil{Kiepenheuer-Institut f\"ur Sonnenphysik, Sch\"oneckstra{\ss}e 6,
D-79104 Freiburg, Germany; and

Department of Physical Sciences, Astronomy Division, P.O. Box 3000,
FIN-90014 University of Oulu, Finland}

\newcommand{\EQ}{\begin{equation}}
\newcommand{\EN}{\end{equation}}
\newcommand{\EQA}{\begin{eqnarray}}
\newcommand{\ENA}{\end{eqnarray}}

\newcommand{\Eq}[1]{equation~(\ref{#1})}

\newcommand{\Fig}[1]{Fig.~\ref{#1}}

\newcommand{\Figss}[2]{Figs~\ref{#1}--\ref{#2}}

\newcommand{\bra}[1]{\langle #1\rangle}

{}

\newcommand{\meanAA}{\overline{\bm{A}}}
\newcommand{\meanBB}{\overline{\bm{B}}}
\newcommand{\meanJJ}{\overline{\bm{J}}}
\newcommand{\meanUU}{\overline{\bm{U}}}

\newcommand{\meanuu}{\overline{\mbox{\boldmath $u$}}{}}{}
{}
{}
\newcommand{\meanEMF}{\overline{\mbox{\boldmath ${\cal E}$}}{}}{}
{}
{}
{}
{}
\newcommand{\meanA}{\overline{A}}
\newcommand{\meanB}{\overline{B}}

\newcommand{\meanU}{\overline{U}}

\newcommand{\nnn}{\hat{\mbox{\boldmath $n$}} {}}

\newcommand{\uu}{{\bm{u}}}

\newcommand{\jj}{{\bm{j}}}

\newcommand{\bb}{{\bm{b}}}

\newcommand{\nab}{\mbox{\boldmath $\nabla$} {}}

\newcommand{\oo}{\mbox{\boldmath $\omega$} {}}

\newcommand{\DD}{{\rm D} {}}

\def\onethird{{\textstyle{1\over3}}}

\newcommand{\nHz}{\,{\rm nHz}}

\newcommand{\yapj}[3]{ #1, {ApJ,} {#2}, #3}

\newcommand{\yan}[3]{ #1, {AN,} {#2}, #3}

\newcommand{\yana}[3]{ #1, {A\&A,} {#2}, #3}

\newcommand{\yjfm}[3]{ #1, {JFM,} {#2}, #3}

\newcommand{\yprt}[3]{ #1, {Phys. Rep.,} {#2}, #3}
\newcommand{\yprl}[3]{ #1, {PRL,} {#2}, #3}

\newcommand{\ymn}[3]{ #1, {MNRAS,} {#2}, #3}

\newcommand{\ysph}[3]{ #1, {Solar Phys.,} {#2}, #3}

\newcommand{\yjour}[4]{ #1, {#2}, {#3}, #4}

\newcommand{\sanac}[2]{ #1, {A\&A,} (submitted, #2)}

\begin{document}


\maketitle

\begin{abstract}
A two-dimensional mean field dynamo model is solved where magnetic helicity
conservation is fully included.
The model has a negative radial velocity gradient giving rise to
equatorward migration of magnetic activity patterns.
In addition the model develops longitudinal variability
with activity patches travelling in longitude.
These patches may be associated with active longitudes.
\end{abstract}

\section{Introduction}

Active regions are complexes of magnetic activity out of which
sunspots, flares, coronal mass ejections, and several other phenomena
emerge with some preference over other regions.
These regions tend to be bipolar, i.e.\ they come in pairs of opposite
polarity and are roughly aligned with the east--west direction.

There is some controversy as to whether or not active regions
appear preferentially along so-called active longitudes and what
their long term stability properties are (e.g., Bai 2003,
Berdyugina \& Usoskin 2003, Pelt et al.\ 2005).
Some degree of recurrence of sunspots has frequently been
reported over the years (Vitinskij 1969, Bogart 1982, Bai 1987, 1988),
but only now these ideas are becoming more quantitative.

As the recent analysis of Usoskin et al.\ (2005) has shown,
active longitudes have characteristic angular velocities that
depend on the phase of the cycle which, in turn,
determines the typical latitude of their occurrence.
The analysis of solar magnetograms
by Benevolenskaya et al.\ (1999) showed already that
at the beginning of each cycle, when most of the activity
occurs at about $\pm30^\circ$ latitude, the rotation rate
of the active longitudes is $\Omega/2\pi\approx446\nHz$,
while at the end of each cycle, when the typical latitude is
only $\pm4^\circ$ latitude, the rotation rate
of the active longitudes is $\Omega/2\pi\approx462\nHz$.
The recent work of Usoskin et al.\ (2005) demonstrates that the
active regions can also be detected in sunspot data.
Unlike the analysis of Benevolenskaya et al.\ (1999),
who isolated two different active longitudes at the beginning
and the end of the cycle, Usoskin et al.\ (2005) determined
a continuous latitudinal dependence of these active longitudes
on the phase of the cycle.
The success of their analysis lies in the way they calculated
time-dependent reference values.
In particular, they find that only 10\% of the spots participate in this
nonaxisymmetric effect.
Thus, the effect is real, but weak, although with a well determined strength.

Given that active longitudes suggest the presence of well preserved
activity patches in an otherwise turbulent medium, one has to look
for a quantity  that has the capability to be long-lived.
An obvious candidate for anything long lived in hydromagnetic turbulence
at large magnetic Reynolds numbers is the magnetic helicity.
In the absence of magnetic helicity fluxes, magnetic helicity is nearly
perfectly conserved in resistive magnetohydrodynamics at large
magnetic Reynolds numbers.
Even in the presence of magnetic helicity fluxes, magnetic helicity is
only transferred from one place to another, but it is not lost.
It is therefore plausible that there might be a connection between the
nearly perfect magnetic helicity conservation and the long-lived features
on the sun such as active longitudes.
In the simplest case, magnetic helicity could be considered frozen into
the plasma so that local patches of enhanced magnetic helicity would just
propagate with the ambient velocity of the gas.
This could in principle be a very simplistic picture of active longitudes
that might explain their long life times, but not how they came into
existence.

An alternative that is traditionally discussed in connection with
active longitudes is the idea that there are
a number of different axisymmetric and nonaxisymmetric dynamo modes
present in the sun that are mixed in the right proportions such that their
superposition corresponds to the observed field configuration
(R\"adler et al.\ 1990, Moss 1999, 2004, Moss \& Brooke 2000,
Bigazzi \& Ruzmaikin 2004, Berdyugina et al.\ 2006).
An obvious problem is that these modes are solutions of the linearized
problem and that their superposition does not constitute a solution to
the nonlinear problem.
At first instance only one of the modes gets selected, so the final solution
is sill mainly either axisymmetric or nonaxisymmetric, but not easily
anything in between, as originally anticipated (R\"adler et al.\ 1990,
Moss et al.\ 1995).

Magnetic helicity conservation only applies to the total field, i.e.\
the sum of small scale and large scale fields.
The large scale field produced by a mean field (large scale) dynamo
does by itself not conserve magnetic helicity, because the $\alpha$ effect
leads to a transfer of magnetic helicity from smaller to larger scales
(Pouquet et al.\ 1976, Ji 1999).
It is therefore necessary to include also the contribution from the small
scale magnetic helicity, which enters into the mean field description
through a magnetic contribution to the $\alpha$ effect.
This approach is now fairly well developed and is important in
reproducing the slow saturation (Field \& Blackman 2002,
Blackman \& Brandenburg 2002, Subramanian 2002) found in simulations
(Brandenburg 2001).
For a review of these recent developments see
Brandenburg \& Subramanian (2005a).

We begin by explaining this approach in more detail.
In order to focus on the essentials, we consider only a minimalistic
model.
The basic dynamo wave can already be described in a one-dimensional
model with only latitudinal extent.
In the present context we still need the longitudinal extent, but we
ignore the variation of the magnetic field with depth.
Furthermore, in order to study the basic effect introduced by
the longitudinal extent and by magnetic helicity conservation, it
suffices to restrict oneself to a cartesian model.
Since there is no radial extent, there are also no radial
boundaries, and hence no magnetic helicity flux out of them.
This might be an important limitation to keep in mind.

\section{The model}

The dynamo equation together with the dynamical quenching model is
(Field \& Blackman 2002, Blackman \& Brandenburg 2002, Subramanian 2002)
\EQ
{\partial\meanBB\over\partial t}=
\nab\times(\meanUU\times\meanBB+\meanEMF-\eta\mu_0\meanJJ),
\label{fullset1flux}
\EN
\begin{equation}
{\DD\alpha_{\rm M}\over\DD t}=-2\eta_{\rm t} k_{\rm f}^2\left(
{\meanEMF\cdot\meanBB\over B_{\rm eq}^2}
+{\alpha_{\rm M}\over R_{\rm m}}\right),
\label{fullset2flux}
\end{equation}
where $\meanBB$ is the mean magnetic field,
$\meanJJ=\nab\times\meanBB/\mu_0$ is the mean current density,
$\mu_0$ is the vacuum permeability,
$\eta$ is the microscopic magnetic Spitzer diffusivity,
$\eta_{\rm t}$ is the turbulent magnetic diffusivity,
\EQ
R_{\rm m}=\eta_{\rm t}/\eta
\EN
is here used as the definition of the magnetic Reynolds number,
$k_{\rm f}$ is the wavenumber of the energy-carrying scale,
$\DD/\DD t=\partial/\partial t+\meanUU\cdot\nab$ is the advective
derivative with respect to the mean flow $\meanUU$, and
\EQ
\meanEMF=q(\alpha\meanBB -\eta_{\rm t}\mu_0\meanJJ)
\label{meanEMFdef}
\EN
is the turbulent electromotive force (assuming local isotropy),
with an {\it ad hoc} added overall quenching of the electromotive force
via the factor
\EQ
q=1/(1+q_B\meanBB^2/B_{\rm eq}^2).
\label{quenching}
\EN
Such an overall quenching was applied also in the recent dynamically
quenched mean field of Brandenburg \& Subramanian (2005b).

Under the assumption of locally isotropic turbulence we have
$\alpha = \alpha_{\rm K}+\alpha_{\rm M}$, where
$\alpha_{\rm K} = -\frac13\tau\overline{\oo\cdot\uu}$
is the kinetic $\alpha$ effect,
where $\oo=\nab\times\uu$ is the vorticity,
$\alpha_{\rm M} = \frac13\tau\overline{\jj\cdot\bb}/\rho_0$ is the
magnetic contribution to the $\alpha$ effect, and $\eta_{\rm t}
=\frac13\tau\overline{\uu^2}$ is turbulent magnetic diffusivity.
In deriving \Eq{fullset2flux} we have used
the definitions $B_{\rm eq}^2=\mu_0\rho \overline{u^2}$ and
$\eta_{\rm t}=\onethird\tau \overline{u^2}$, so that
$\frac13\tau/(\mu_0\rho)=B_{\rm eq}^2/\eta_{\rm t}$
(see Blackman \& Brandenburg 2002).

Equation~(\ref{fullset2flux}) can now be solved together with
\Eq{fullset1flux} to determine
the effect of magnetic helicity evolution on the solar dynamo.
These equations have been solved in recent years both in context
of the galactic dynamo (Kleeorin et al.\ 2000, 2002, Shukurov et al.\ 2005)
and the solar dynamo (Kleeorin et al.\ 2003, Zhang et al.\ 2006).
The dynamical quenching model has also been tested against direct
simulations (Blackman \& Brandenburg 2002, Brandenburg \& Subramanian 2005b).

We emphasize however that, although $\meanBB$ depends only on $x$ and $y$,
$\meanUU$ depends also on $z$ in a prescribed fashion, such as to model
the radial differential rotation.
To ensure that the field stays numerically
divergence free, it is advantageous to write
\Eq{fullset1flux} in terms of the mean magnetic vector potential $\meanAA$,
so $\meanBB=\nab\times\meanAA$ and the radial velocity gradient is explicitly
added to the model by working with the velocity in the form
\EQ
\meanUU=\meanUU^{(0)}(z)+\meanuu(x,y,t);
\quad\meanUU^{(0)}=(Sz,0,0).
\EN
We apply the equations at the reference height $z=0$, so $\meanUU^{(0)}$
itself is zero and does not enter the equations; only a single component
of its derivative matrix, $\meanUU^{(0)}_{x,z}=S$, enters.
Thus, the full set of equations is
\EQ
{\DD\meanA_i\over\DD t}=-\meanU_{j,i}\meanA_j+\meanEMF+\eta\nabla^2\meanA_i,
\EN
\EQ 
{\DD\meanU_i\over\DD t}=-\meanU_{i,j}\meanU_j-c_{\rm s}^2\nabla_i\ln\rho
+\rho^{-1}(\meanJJ\times\meanBB)_i,
\EN
\EQ
{\DD\ln\rho\over\DD t}=-\nab\cdot\meanUU,
\EN
\begin{equation}
{\DD\alpha_{\rm M}\over\DD t}=-2\eta_{\rm t} k_{\rm f}^2\left(
{\meanEMF\cdot\meanBB\over B_{\rm eq}^2}
+{\alpha_{\rm M}\over R_{\rm M}}\right).
\label{fullset2flux2}
\end{equation}
We model the magnetic field only in one hemisphere and restrict ourselves
to dipolar parity by demanding
\EQ
\meanA_{x,y}=\meanA_{y}=\meanA_{z,y}=0\quad\mbox{(on $y=0$)},
\EN
corresponding to a normal-field condition ($\nnn\times\meanBB=0$),
modeling the equator, and
\EQ
\meanA_{x}=\meanA_{y,y}=\meanA_{z}=0\quad\mbox{(on $y=2\pi$)},
\EN
corresponding to a perfect conductor boundary condition
($\nnn\cdot\meanBB=0$), in an attempt to capture some of the behavior
on the pole.
The simulations have been carried out using the {\sc Pencil Code}\footnote{
\url{http://www.nordita.dk/software/pencil-code}} which is a high-order
finite-difference code (sixth order in space and third
order in time) for solving the compressible hydromagnetic equations.
The code comes with a special mean field module for the dynamical
$\alpha$ quenching equations.

\begin{figure}[t!]\begin{center}
\includegraphics[width=\columnwidth]{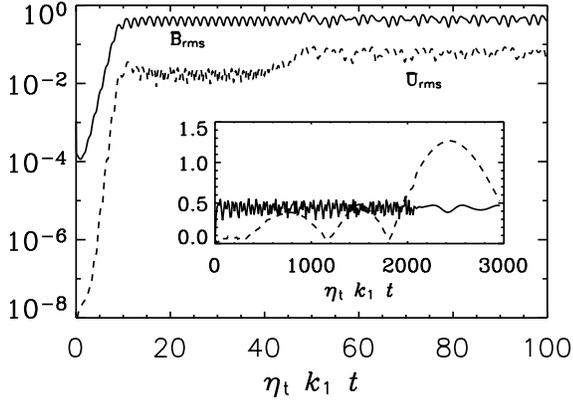}
\end{center}\caption[]{
Time dependence of the rms mean magnetic field and the
rms mean velocity, showing the emergence of oscillatory
and later more irregular solutions.
}\label{pn_B32e_again}\end{figure}

\begin{figure}[t!]\begin{center}
\includegraphics[width=\columnwidth]{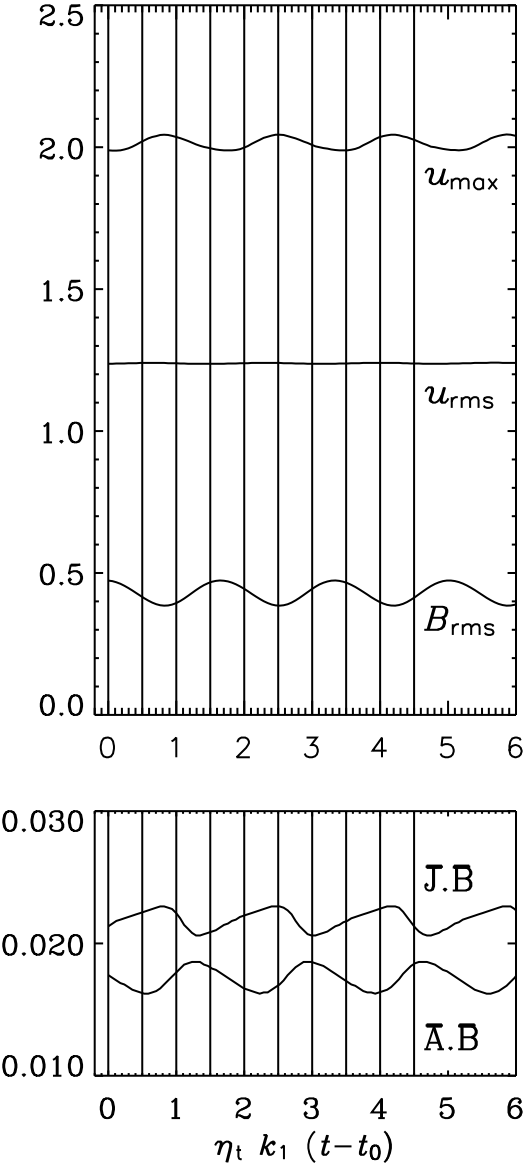}
\end{center}\caption[]{
Time dependence of various integral quantities, in the quiescent
phase after $\eta_{\rm t}k_1t_0=2700$.
The vertical bars indicate the times for which synthetic `magnetograms'
and plots of other quantities are given below.
}\label{pn}\end{figure}

\begin{figure}[t!]\begin{center}
\includegraphics[width=\columnwidth]{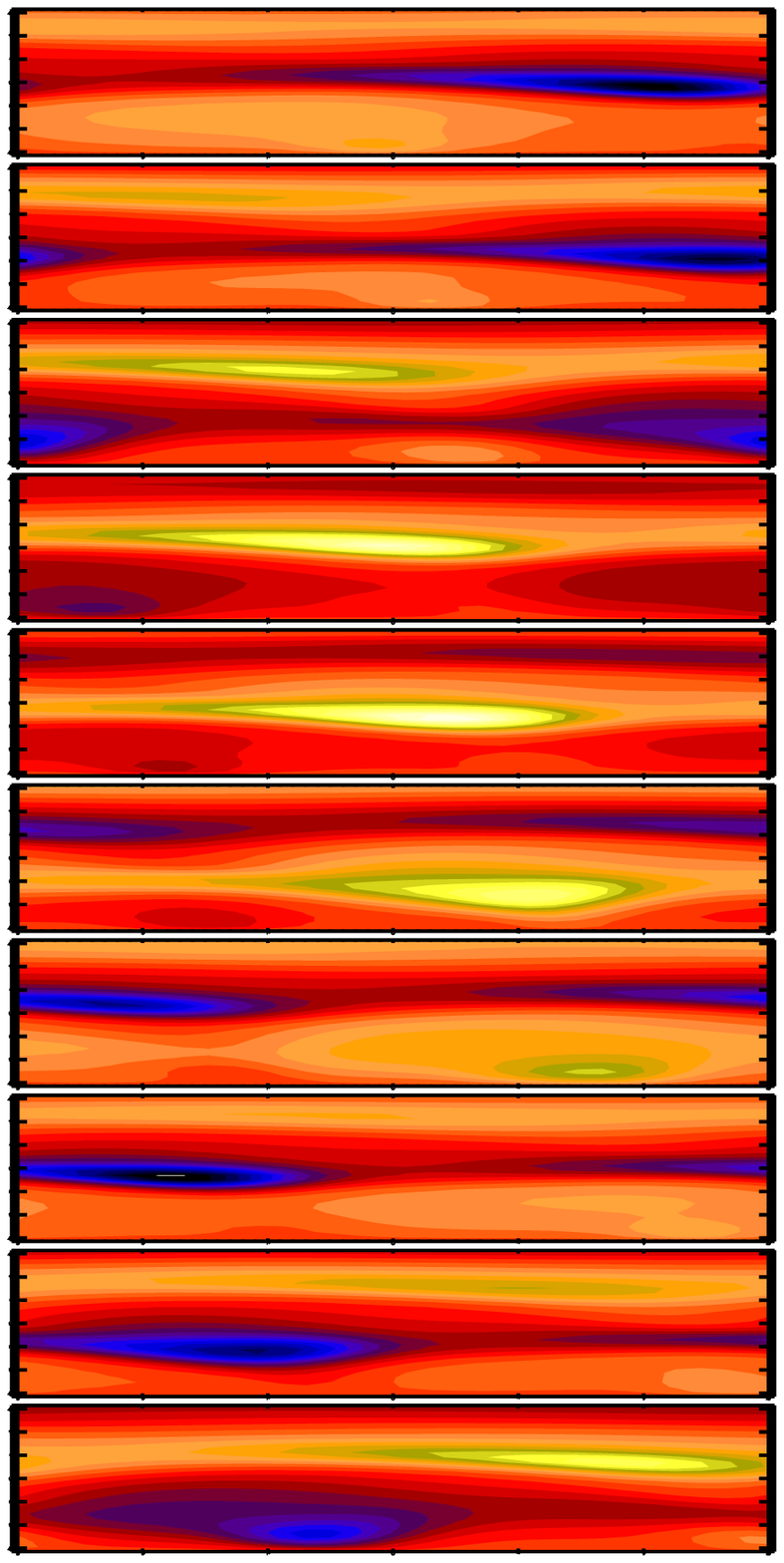}
\end{center}\caption[]{
Plot of $\meanB_z(x,y)$ at different times,
in regular intervals of $\eta_{\rm t}k_1\Delta t=0.5$ after $t=t_0$.
}\label{pbb_all}\end{figure}

\begin{figure}[t!]\begin{center}
\includegraphics[width=\columnwidth]{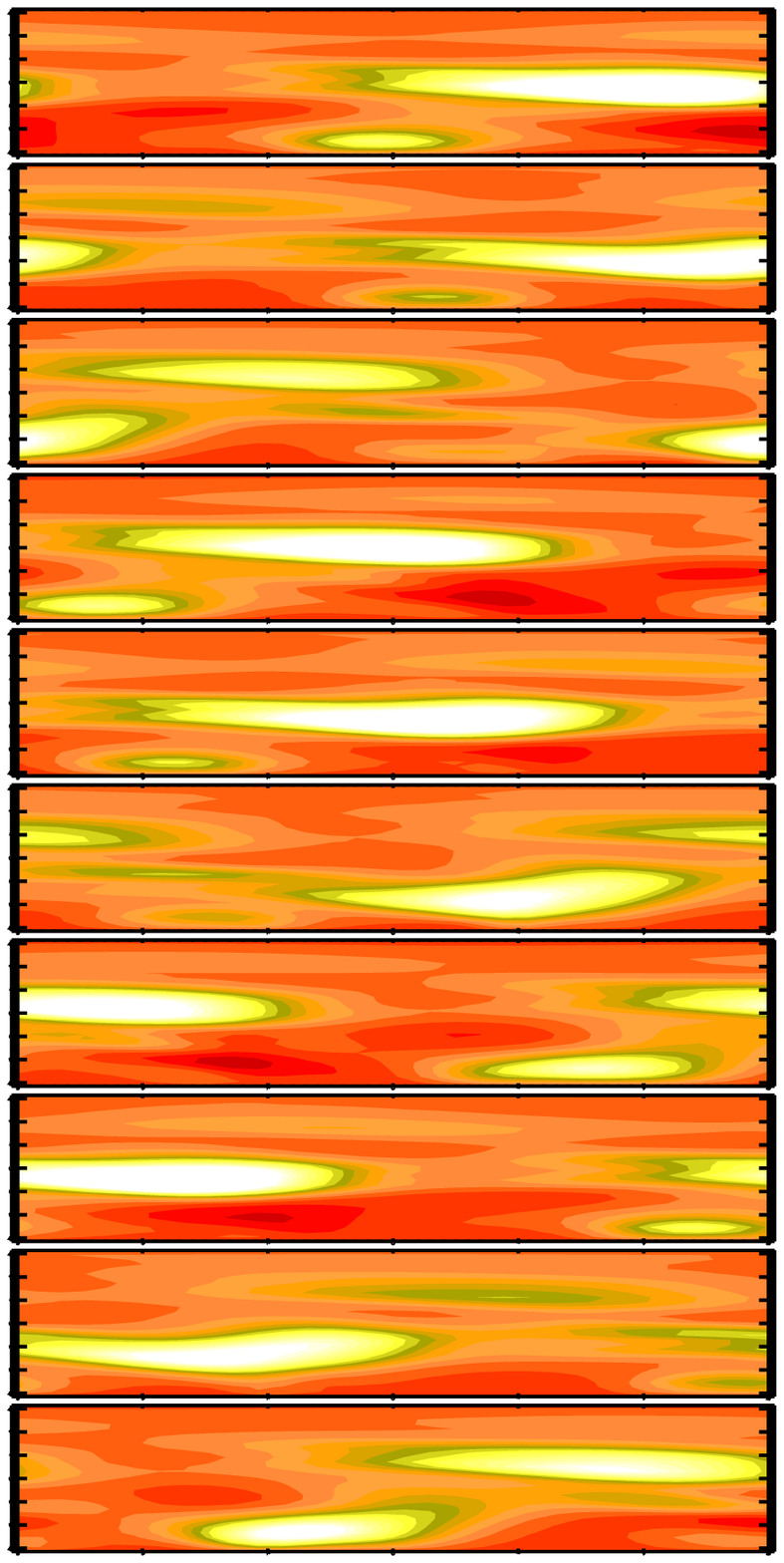}
\end{center}\caption[]{
Plot of $\meanJJ\cdot\meanBB$ at different times,
in regular intervals of $\eta_{\rm t}k_1\Delta t=0.5$ after $t=t_0$.
}\label{pjb_all}\end{figure}

\begin{figure}[t!]\begin{center}
\includegraphics[width=\columnwidth]{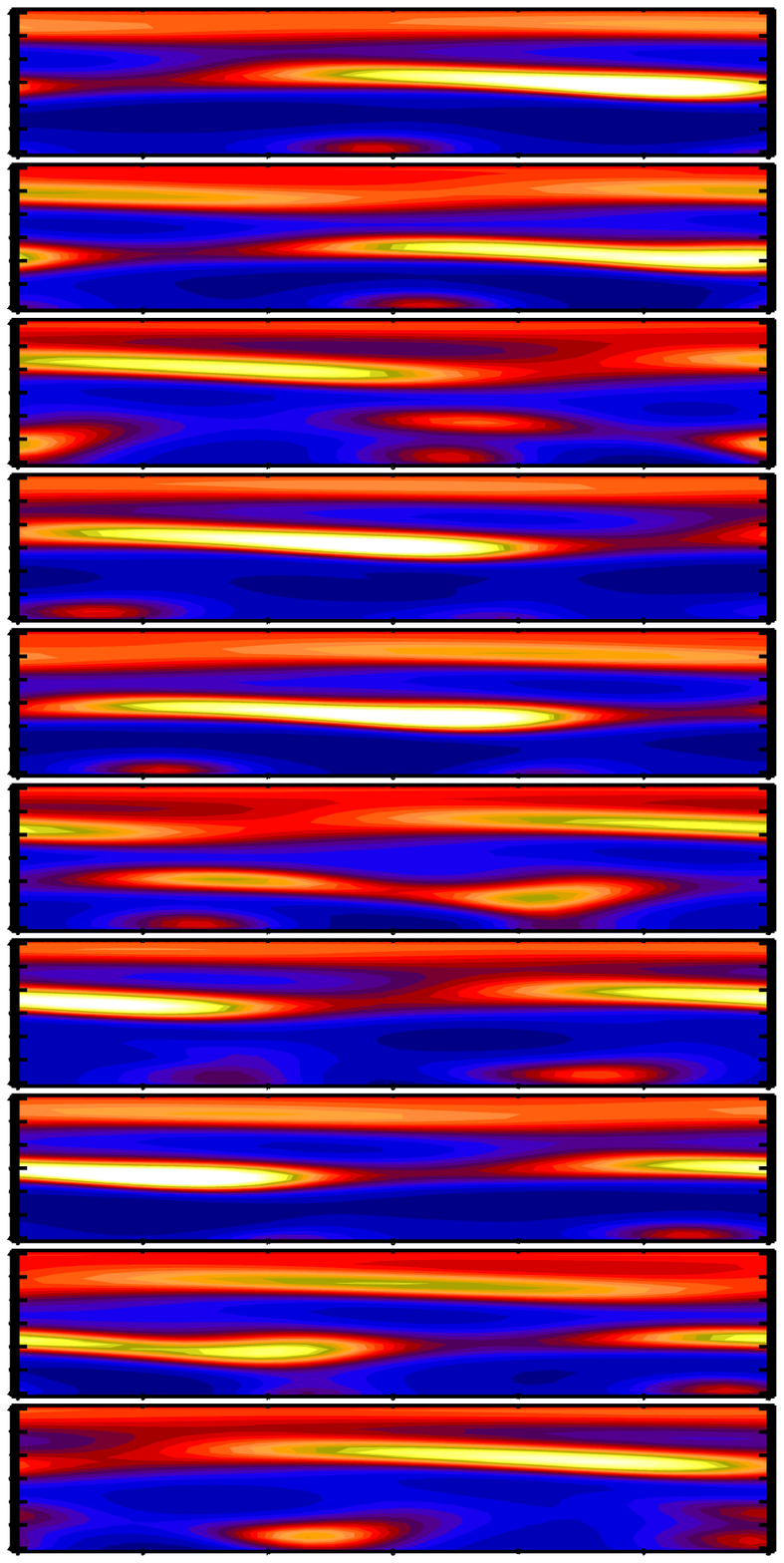}
\end{center}\caption[]{
Plot of $\alpha_{\rm M}(x,y)$ at different times,
in regular intervals of $\eta_{\rm t}k_1\Delta t=0.5$ after $t=t_0$.
}\label{palpm_all}\end{figure}

\section{Results}

In the following we discuss in detail one particular model
using the following set of parameters: $S=-30$ (corresponding
to negative radial differential rotation), $\alpha_{\rm K}=0.5$,
$\nu_{\rm t}=\eta_{\rm t}=1$, $\eta=0.05$ (so $R_{\rm m}=20$),
$k_{\rm f}=5$, and an overall quenching factor with $q_B=0.1$.

In \Fig{pn_B32e_again} we show the evolution of the rms values
of the mean magnetic field, $\bra{\meanBB^2}^{1/2}$ and
of the mean velocity field, $\bra{\meanUU^2}^{1/2}$.
Initially, the magnetic field grows exponentially, but then it
saturates in an oscillatory fashion.
The mean field also drives a mean flow.
The long time behavior is more complicated, as is shown in the
inset: the velocity increases slightly, but then the field
becomes more irregular.
There are also stages where the velocity exceeds the magnetic
field and limits the time dependence.

In \Fig{pn} we show the evolution of the rms values of velocity and
magnetic field, as well as the maximum velocity, together with the
mean magnetic and current helicities, $\bra{\meanAA\cdot\meanBB}$ and
$\bra{\meanJJ\cdot\meanBB}$ during a short time interval during
a quiescent stage.
The vertical bars mark the times for which snapshots of various
quantities are shown in \Figss{pbb_all}{palpm_all} that will be
discussed next.

We begin with \Fig{pbb_all}, where we show images of the line of sight
component of the magnetic field, $\meanB_z$.
One clearly sees magnetic activity patches moving both equatorward
(downward in the plot) as well as in the prograde direction (to the
right in the plot).
The fact that these patches propagate at all is interesting.
It is not related to a locally enhanced mean flow in the prograde
direction: the flow is actually in the retrograde direction.
The propagation must therefore be related to a proper three-dimensional
dynamo mode travelling to the right, just like in the nonaxisymmetric
dynamo models mentioned in the beginning.

Next, we consider images of current helicity, $\meanJJ\cdot\meanBB$,
in \Fig{pjb_all}, which clearly shows enhanced positive current helicity
in the regions where also the line-of-sight magnetic field is strong.
The current helicity has the same sign in the magnetic patches with
positive and negative net flux.

Finally, we consider images of the magnetic $\alpha$ effect.
The connection between $\alpha_{\rm M}$ and field strength or
current helicity in not very clear.
Indeed, the two are only loosely connected with each other.
The primary reason for having the magnetic $\alpha$ term is to provide
a nonlinear feedback between the produced large scale helicity and
the $\alpha$ effect that helps to maintain total magnetic helicity
conservation.

\section{Conclusions}

The present investigations can only be regarded as preliminary, because
we have not been able to explore the big parameter space given by the
large number of unknowns.
Most worrisome is the restriction to only small values of the magnetic
Reynolds number.
At the moment we are experiencing problems when we try larger values,
suggesting that the problem may not we well-posed for larger values
of $R_{\rm m}$ and may require modifications.
However, this is still surprising, because similar problems have not
been encountered in simpler problems, where however no mean advection
or shear was taken into account (e.g.\ Brandenburg \& Subramanian 2005b).

One surprising aspect emerging from the present simulations is the
fact that magnetic patterns do not move with the local gas velocity.
Instead, the field propagation is governed by the nonaxisymmetric
dynamics of the dynamo modes.
Thus, these simulations would not support the notion of field line anchoring.
This pictures is occasionally used in connection
with sunspot proper motions.
Long before the internal angular velocity was determined via
helioseismology, it was known that sunspots rotate faster than the
surface plasma (Howard et al.\ 1984).
Moreover, young sunspots rotate faster than old sunspots
(Tuominen \& Virtanen 1988, Pulkkinen \& Tuominen 1998).
A common interpretation is that young sunspots are still anchored
at a greater depth than older ones, and that therefore the internal
angular velocity must increase with depth (see also Brandenburg 2005).
This provided also the basis for the classical mean field dynamo
theory  of the solar cycle according to which the radial angular
velocity gradient has to be negative (Parker 1987).

Given the lack of agreement between the speed of magnetic patches
and the local flow speed, one it led to believe that
magnetic structures can therefore not be used
as tracers of the local flow speed.
Alternatively, it is also possible that the tracer properties of the
advected field are fully displayed only in three dimensions,
or at larger magnetic Reynolds numbers.
However, the present results should be interpreted with care, because the
present calculations have only been possible in a rather limited parameter
range and for rather small magnetic Reynolds numbers.
It would be important to test the notion of field line anchoring in
direct simulations of the original non-averaged equations, i.e.\ in
the presence of developed turbulence.

\section*{Acknowledgments}

The Danish Center for Scientific Computing is acknowledged for granting
time on the Horseshoe cluster.


\end{document}